\documentclass[12pt,preprint]{aastex}
\usepackage{graphicx}
\def \kpd{KPD~0005+5106}

\slugcomment{Accepted for publication in the Astrophysical Journal}
\shorttitle{Deposing the Cool Corona of KPD~0005+5106}
\shortauthors{J.J.~Drake \& K.~Werner}
\begin{document}
\title{Deposing the Cool Corona of KPD~0005+5106}
\author{Jeremy J. Drake\altaffilmark{1} and Klaus Werner\altaffilmark{2}}
\affil{$^1$Smithsonian Astrophysical Observatory,
MS-3, \\ 60 Garden Street, \\ Cambridge, MA 02138}
\email{jdrake@cfa.harvard.edu}
\affil{$^2$Institut f\"ur Astronomie und Astrophysik, 
Sand 1, 72076 T\"ubingen, Germany}
\email{werner@astro.uni-tuebingen.de}

\begin{abstract}
The ROSAT PSPC pulse height spectrum of the peculiar He-rich hot white
dwarf KPD~0005+5106 provided a great surprise when first analysed by
Fleming, Werner \& Barstow (1993).  It defied the best non-LTE
modelling attempts in terms of photospheric emission from He-dominated
atmospheres including C, N and O and was instead interpreted as the
first evidence for a coronal plasma around a white dwarf.  We show
here that a recent high resolution {\it Chandra} LETGS spectrum has 
more structure than expected from a thermal bremsstrahlung continuum
and lacks the narrow lines of H-like and He-like C expected from a
coronal plasma.  Moreover, a coronal model requires a total luminosity
more than two orders of magnitude larger than that of the star itself.
Instead, the observed 20-80~\AA\ flux is consistent with photospheric
models containing trace amounts of heavier elements such as Fe.  The
soft X-ray flux is highly sensitive to the adopted metal abundance and
provides a metal abundance diagnostic.  The weak
X-ray emission at 1~keV announced by O'Dwyer et al (2003)
instead cannot arise from the photosphere and requires alternative
explanations.  We echo earlier speculation that such emission arises
in a shocked wind.  Despite the presence of UV-optical O~VIII lines
from transitions between levels $n=7$-10, no X-ray O~VIII
Ly$\alpha$ flux is detected.  We show that O~VIII Lyman photons can be
trapped by resonant scattering within the emitting plasma 
and destroyed by photoelectric 
absorption.
\end{abstract}


\section{Introduction}
\label{s:intro}

The hot $10^6$-$10^7$~K coronae on the Sun and other late-type stars
are believed to be sustained by mechanical energy in their outer
convection zones, which is dissipated at the surface through the
medium of magnetic fields generated and amplified by differential
rotation and convection in the interior.  This paradigm is reinforced
by the decline and disappearance of X-ray emission that occurs toward
hotter stars at spectral type A, which also corresponds to the
disappearance of outer convection zones.  While the extensive {\it
Einstein} stellar survey \citep{Vaiana.etal:81} detected copious
X-rays from earlier O- and B-type stars in addition to later types,
this has generally been attributed to emission from radiatively-driven
supersonic shock-heated winds rather than coronal
plasma \citep{Cassinelli:82}, although the possible contributions from
magnetically-confined 
plasma have recently been reconsidered based on recent {\it Chandra}
and {\it XMM-Newton} high resolution spectra
\citep[e.g.][]{Miller:02}.

In the above context, the low resolution {\it ROSAT}
Position-Sensitive Proportional Counter (PSPC) pulse height spectrum
of the unusual hot He-rich white dwarf \kpd\
\citep[WD~0005+511]{Downes.etal:85} presented a
major surprise.  \citet{Fleming.etal:93} found that they could not
match this spectrum using the best available spectral models computed 
using non-LTE model atmospheres containing He, C, N and O.  
Instead, they found that a
satisfactory fit to the data could be obtained from thermal
bremsstrahlung and optically-thin plasma models with temperatures of
$2-3\times 10^5$~K.  The apparent impropriety of photospheric spectral
models and success of thermal plasma models provided evidence that the
X-rays originated not from deeper atmospheric layers but from a {\em
coronal plasma} encircling the star.  \kpd\ then became the first
white dwarf thought to have a corona, albeit a cool one compared to
the $10^6-10^7$~K coronae of late-type stars.  As
\citet{Fleming.etal:93} noted, this result was quite novel and
surprising, since all previously observed X-ray emission from white
dwarfs had been photospheric in nature.  Moreover, while DA and DB
stars can possess significant convection zones corresponding to H and
He ionization layers and might conceivably sustain magnetic dynamo
activity, this is not the case for the much hotter $1.2\times 10^5$~K
effective temperature of \kpd\ \citep{Werner.etal:94} where He is
completely ionized.

A corona about \kpd\ is perhaps a less surprising proposition when
viewed in the light of other peculiarities of its spectrum.  Emission
lines of C~V, N~V and O~VIII have been detected at UV and optical
wavelengths, prompting the suggestion of ongoing, and possibly
shock-heated, mass loss
\citep{Werner.etal:94,Sion.Downes:92,Sion.etal:97}.  More recently,
\citet{O'Dwyer.etal:03} and \citep[see also][]{Chu.etal:04b} have
reported a detection of weak X-ray emission at 1~keV based on a
re-analysis of deeper ROSAT PSPC observations of \kpd , while
\citet{Otte.etal:04} have presented the discovery of an O~VI-emitting
nebula associated with the object.

In this paper, however, we show that the {\it Chandra} Low Energy
Transmission Grating Spectrograph and High Resolution Camera
Spectroscopic detector (LETG+HRC-S) spectrum of \kpd\ can be
qualitatively modelled as photospheric emission, and does not require
a coronal explanation.  The excess flux found by
\citet{Fleming.etal:93} in the 0.2-0.3~keV range arises from deeper
atmospheric layers revealed by a lower continuous opacity in this
region.  Difficulties with earlier photospheric modelling attempts are
shown to have arisen because of the neglect of trace heavier elements
such as Fe.  The 1~keV emission found by \citet{O'Dwyer.etal:03} lies
below our detection threshold, but cannot be explained by photospheric
models.

\section{Observations and Analysis}
\label{s:obs}


\kpd\ was observed by {\it Chandra} using the 
LETG+HRC-S in its standard configuration on 2000 October 25 between UT
05:16 and 10:58 for a total of 19114s, after correction for instrument
deadtime and bad event filtering.

\begin{figure}
\epsscale{1.0}
\plotone{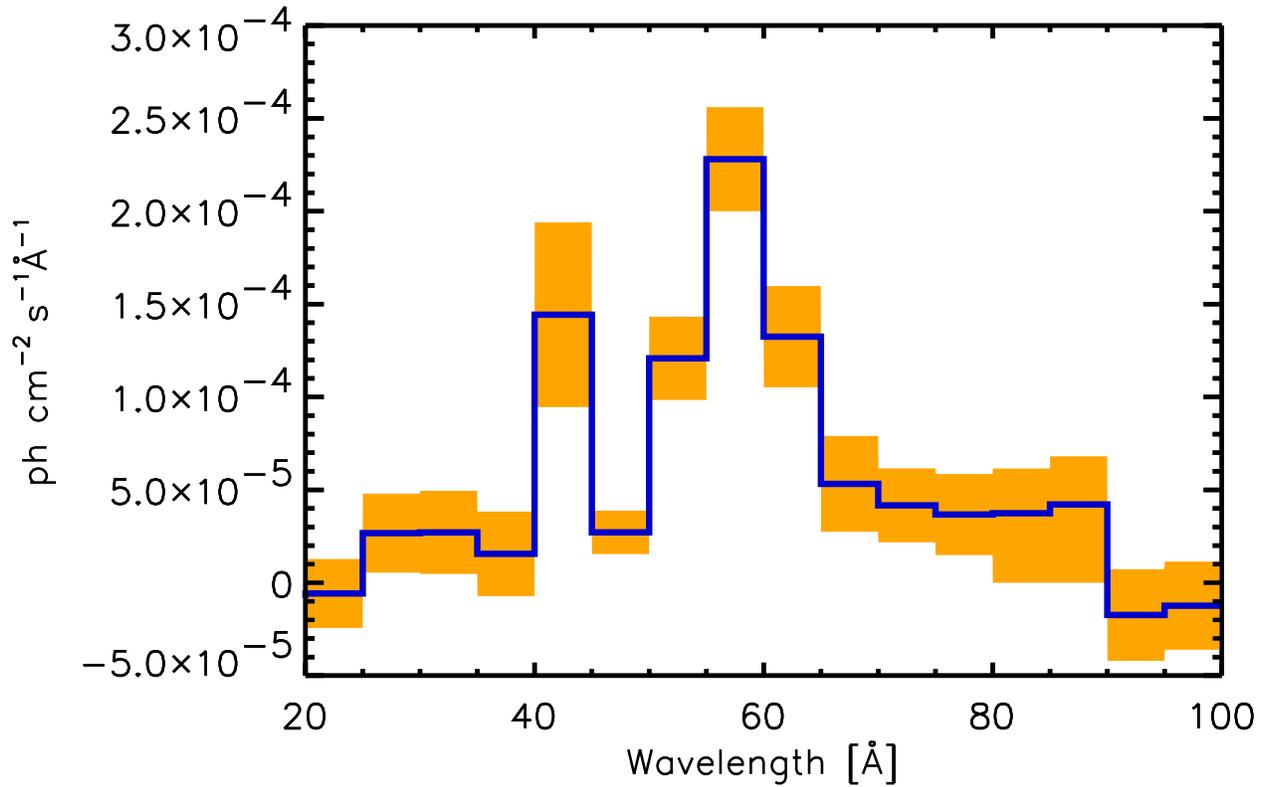}
\caption{The {\it Chandra} LETG+HRC-S spectrum of \kpd ; shaded regions
represent $1\sigma$ uncertainties.  No evidence for narrow
lines is present in the data.  We therefore show here the spectrum binned
at 5~\AA\ intervals in order to illustrate the observed continuum
structure.}
\label{f:spec}
\end{figure}

Initial reduction of satellite telemetry was performed by the {\it
Chandra} X-ray Center Standard Data Processing software, but was
completely reprocessed by us using CIAO 3.1.  The analysis described
here is based on the Level 2 products of this reprocessing.  The
scientific analysis was undertaken using the PINTofALE\footnote{Freely
available from http://hea-www.harvard.edu/PINTofALE/}
IDL\footnote{Interactive Data Language, Research Systems Inc.}
software suite \citep{Kashyap.Drake:00}.  The spectrum of \kpd\ was
extracted by summing the events within a window $3.6\arcsec$ in width
in the cross-dispersion direction, and background was estimated in two
strips of width $18\arcsec$ each located on either side of the
spectral trace.  The final extracted spectrum, binned at 5~\AA\
intervals, is illustrated in
Figure~\ref{f:spec}.  The distribution of
the dispersed photon events is consistent with a continuum source: no
narrow emission line features were discernible.

Our analysis involved comparison of the observed LETGS spectrum of
\kpd\ with the emergent spectra computed for a range of line blanketed
non-LTE model atmospheres and for optically-thin collision-dominated
coronal models.  Coronal radiative loss models were computed using
PINTofALE, employing line and continuum emissivities
computed using atomic data from the CHIANTI database version 4.02
\citep{Young.etal:03} and the ionization equilibrium of
\citet{Mazzotta.etal:98}.

Model atmosphere computations were performed using the PRO2
code \citep{Werner.etal:03} assuming hydrostatic and radiative
equilibrium.  Model atoms and atomic data used for
detailed non-LTE calculations have been described recently by
\citet{Werner.etal:04}.  Constraints for the input model parameters
are provided by earlier analyses of UV spectra obtained by the HST
Faint Object Spectrograph (FOS) and Goddard High Resolution
Spectrograph (GHRS), and optical spectra obtained at the 3.5~m
telescope at Calar Alto \citep{Werner.etal:94,Werner.etal:96}.
Additionally, we have examined more recent Far Ultraviolet
Spectroscopic Explorer (FUSE) spectra of \kpd\ to provide further
constraints on the abundances of the elements O and Fe that exhibit lines in
the FUSE bandpass. From the lack of Fe lines an upper abundance limit was
determined by \citet{Miksa.etal:02}. From O~VI lines we have determined the O
abundance with the models computed for the present paper.
Our adopted nominal parameters for \kpd\ are
listed in Table~1.
For a single test calculation including sulfur
we assume a solar S abundance (by mass fraction). The absolute flux
calibration for our 
models is provided by HST UV observations to high precision, 
and for comparisons with
{\it Chandra} spectra we normalise the models to the observed flux
density of $4.0\times 10^{-12}$~erg~cm$^{-2}$~s$^{-1}$~\AA$^{-1}$ at
1200~\AA.

\begin{table}
\label{t:params}
\begin{center}
\caption{Fundamental Parameters and Abundances for \protect{\kpd}}
\begin{tabular}{lc}
\hline
$T_{\rm eff}$ & $120000 \pm 10000$ K \\
$\log g$ & $7.0 \pm 0.5$ \\
H/He 	& $\leq -0.7$ \\
C/He\tablenotemark{a} & $-3 \pm 0.5$ \\
O/He	& $-3.8\pm 0.5$ \\
Ne/He   & $\leq -8$ \\
S/He    & $-4.4$ \\
Fe/He	& $\leq-5$   \\ 
$N_H$	& $(5\pm 1.1) \times 10^{20}$~cm$^{-2}$ \\ 
$F_{1200}$\tablenotemark{b} & $4.0\times 10^{-12}$ \\ \hline
\end{tabular}
\tablenotetext{a}{X/He$=\log_{10} n(X)/n(He)$}
\tablenotetext{b}{Flux density at 1200~\AA\ in units of
erg~cm$^{-2}$~s$^{-1}$~\AA$^{-1}$} 
\end{center}
\end{table}


The reasoning of \citet{Fleming.etal:93} leading to the conclusion
that \kpd\ has a corona was based on the comparison of low resolution
X-ray pulse height spectra with model predictions for atmospheres
containing C, N and O.  We argue in this paper that difficulties in
matching the X-ray data with photospheric models was primarily due to the
neglect of heavier elements that can contaminate the photosphere.
While settling under the strong gravity of the white
dwarf tends to empty the atmosphere of heavier elements, the
atmospheres of hotter stars can be substantially enriched by radiative
levitation \citep[e.g.][]{Chayer.etal:95,Schuh.etal:02}.

\begin{figure}
\epsscale{1.0}
\plotone{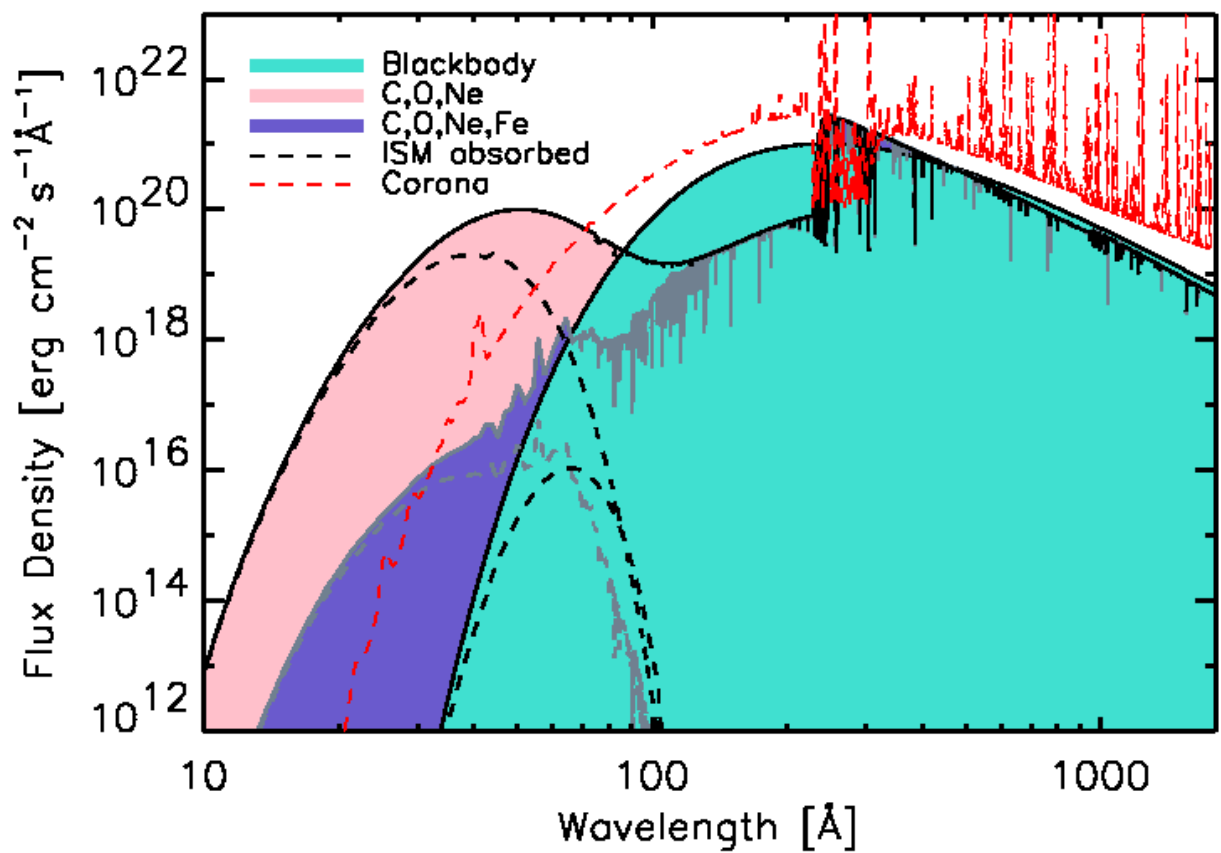}
\caption{Emergent spectra in the UV-X-ray range for non-LTE 
model atmospheres computing using the PRO2 code \citep{Werner.etal:03} for the
parameters and abundances listed in Table~1.
Dashed curves illustrate the spectra absorbed by an intervening
ISM column of $4\times 10^{20}$~cm$^{-2}$.  The model including Fe
assumed Fe/He$=-5$, consistent with the current upper
limit based on UV and optical spectra.  Also shown is a coronal model
computed for a temperature $T=2\times 10^5$~K, normalised to the
{\em relative} level required to explain the ROSAT observations
presented by \citet{Fleming.etal:93}.}
\label{f:fluxes}
\end{figure}

Model fluxes for atmospheres with an effective temperature
$T_{eff}=120000$~K, with and without Fe, are compared in
Figure~\ref{f:fluxes}.  Models include full non-LTE treatment of
C, O, Ne, S and Fe, whose abundances were adopted from Table~1.
Also illustrated is the blackbody flux density
distribution for the same temperature.  It is interesting to note that
the blackbody has much less short wavelength flux than do the more
realistic models: the short wavelength extension of the spectra of the
latter arises because of changes in the opacity with wavelength that
expose the deeper, hotter atmospheric layers at these wavelengths.
Addition of Fe with an abundance equal to the best current upper limit
(Table~1)
however, dramatically reduces the soft X-ray
flux: clearly any photospheric soft X-ray emission from \kpd\ and
similar stars is critically dependent on the photospheric abundances
of heavier elements, as was also demonstrated earlier in the case of
EUV fluxes at longer wavelengths for slightly cooler stars
\citep[e.g.][]{Barstow.etal:95}.

\begin{figure}
\epsscale{0.7}
\plotone{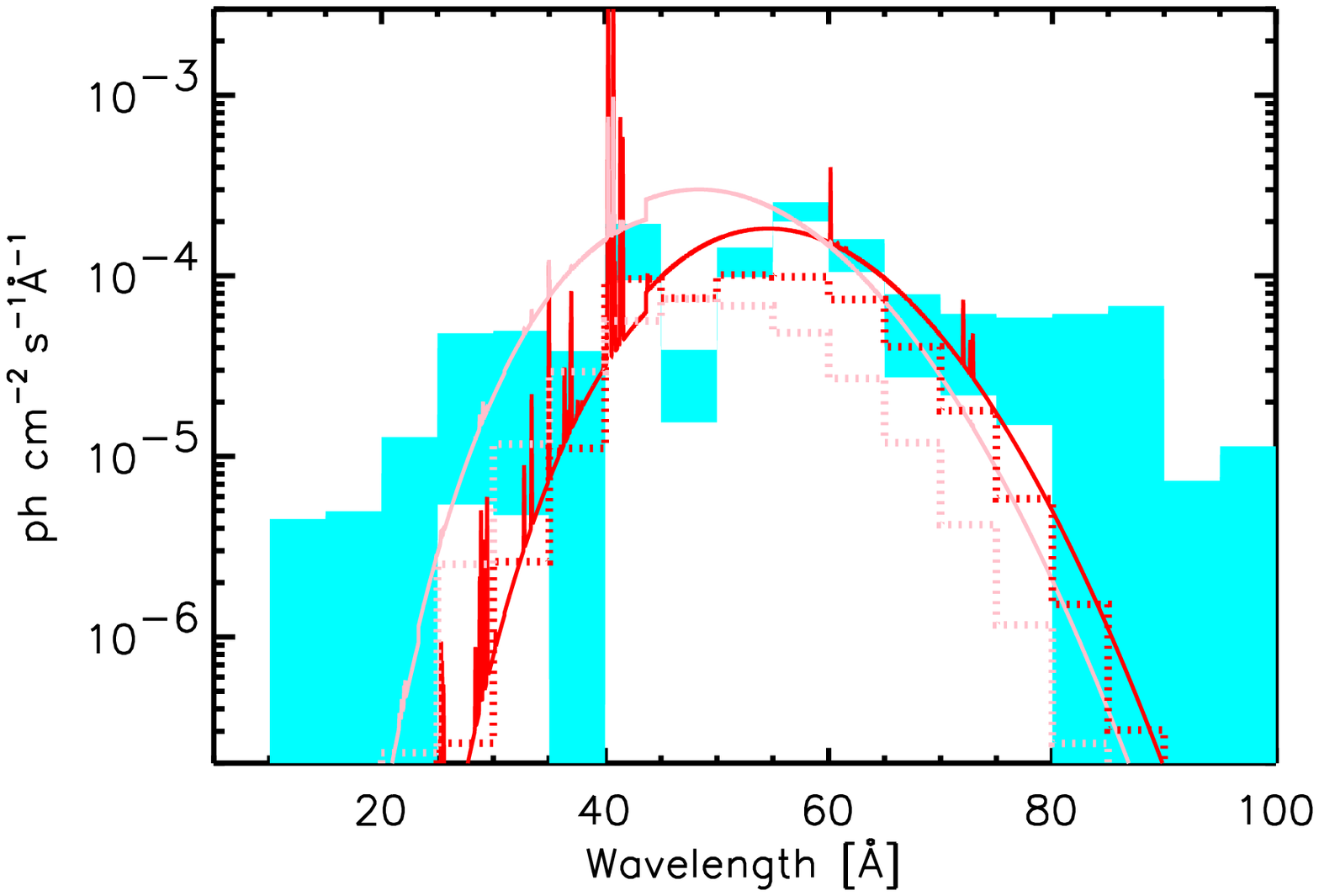}
\plotone{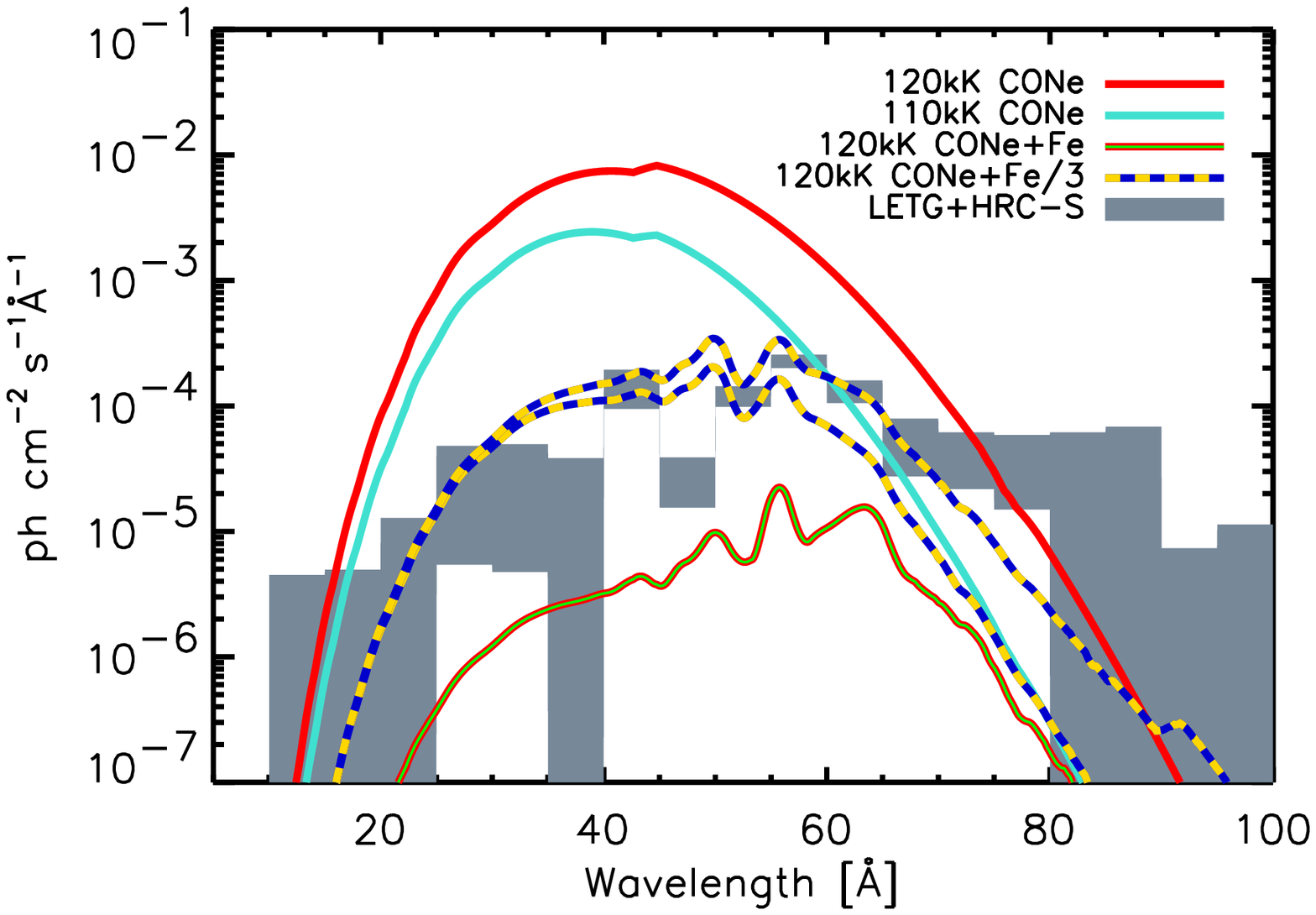}
\caption{Comparison of the observed LETGS spectra (shaded region) with
different coronal (upper) and non-LTE photospheric model (lower)
predictions.  Coronal models are shown both at the theoretical
resolution of the LETGS, and binned at the same 5~\AA\ intervals as
the observations.  Models are attenuated by an ISM absorbing column of
$N_H=4\times 10^{20}$.  Also shown in the lower panel is the ``best fit''
model (``120kK CONe+Fe/3'') for which $N_H=3\times 10^{20}$.
}
\label{f:compare}
\end{figure}

Coronal and photospheric model spectra are compared to the
observations in Figure~\ref{f:compare}.  Coronal models computed for
the He-rich composition listed in Table~1 
are dominated
by the He bound-free and free-free continua, and by the lines of
He-like C near 
40~\AA.  While it is tempting to interpret the peak in the observed
flux between 40 and 45~\AA\ as being due to these C~V lines, there are
no such lines visible in the spectrum where the observed counts are
more smoothly distributed.  The absence of C~V lines does not
necessarily argue against the coronal interpretation of the spectrum,
however, since one might expect heavier elements to gravitationally
settle out of such a plasma.  We therefore also computed models in
which the metal abundances were reduced by a factor of 100.

We confirm the result of \citet{Fleming.etal:93} that the best-fit
coronal spectrum has a temperature in the range 2-$3\times 10^5$~K;
our best-fit model corresponds to $T=2\times 10^5$~K, with an
interstellar absorption component represented by a column of neutral
hydrogen of $3.5\times 10^{20}$~cm$^{-2}$.  Qualitative comparison of
the best-fit coronal model and data suggest, however, that the latter
have a more complex shape than can be achieved by the former.

Photospheric models both with and without Fe were computed, as was a
test model including S; a small 
subsection of the models investigated are illustrated in the lower
panel of Figure~\ref{f:compare}.  It is clear that models with only C,
O and Ne  
greatly overpredict the observed X-ray flux by factors of 10 or
more.  Based on Figure~\ref{f:fluxes}, we see that the emergent soft
X-ray flux can be reduced by addition of Fe and the opacity this element
provides at these wavelengths.  This is borne out in practice, and 
the photospheric spectrum that was found to best match the
observations corresponds roughly to the parameters
$T_{eff}=120000$~K and Fe/He$=-5.5$, with a neutral hydrogen column
density of $3\times 10^{20}$~cm$^{-2}$.  It is clear from
Figure~\ref{f:compare} that there is no obvious reason to prefer
coronal models over photospheric ones.  

While we have not succeeded in producing a photospheric model that
perfectly matches the observations, we have achieved a good
qualitative match by adding only plausible amounts of Fe to the C, O
and Ne abundance mixture.  Trace amounts of other metals such as Na,
Mg, S, Ar, Ca and Ni could also affect significantly the soft X-ray
spectrum, and we could doubtless obtain a better match through their
addition and subsequent optimisation of other parameters such as
effective temperature, surface gravity and ISM absorption.  The test
model including S, for example, introduced an absorption edge near
43~\AA\ with little effect on the emergent flux at longer wavelengths;
careful adjustment of the S abundance could aid in a better prediction
of the observed drop in flux between 40 and 45~\AA .  Further progress
in this direction would demand some additional constraints on the
abundances of these other metals in order to limit the available parameter
space to tractible proportions.

\section{Discussion and Conclusions}
\label{s:discuss}

\subsection{Photospheric vs Coronal Emission}

We have shown that the {\it Chandra} LETG+HRC-S spectrum of \kpd\ in
the 25-90~\AA\ range can be qualitatively modelled by photospheric
emission from an atmosphere containing He, C, O and Ne in expected
amounts, with the addition of Fe at a level consistent with the
current observational upper limit.  With these quite plausible models,
there is no requirement in the data to resort to more exotic solutions
such as a coronal model.  The quality of
the current short LETGS exposure does not allow us to rule out a
significant amount of coronal emission with a high degree of
confidence, though the best fit model based on {\em purely} coronal
emission is statistically unacceptable with a reduced $\chi^2=2.1$.

We can, however, completely discount a coronal interpretation of the ROSAT
and {\it Chandra} spectra on energetic grounds.  A coronal plasma
model at a temperature of 2-$3\times 10^5$~K absorbed by an ISM column
of $N_H\sim 10^{20}$~cm$^{-2}$ and normalised to the observed ROSAT
spectrum has a total luminosity 2-3 orders of magnitude {\em greater} than
than the star itself.  This is qualitatively illustrated in
Figure~\ref{f:fluxes}, where the coronal model is shown normalised
relative to the photospheric models to the approximate {\em relative} 
level required to explain the observed 20-80~\AA\ X-ray flux.

\subsection{Remaining Puzzles}

Despite the ability of photospheric models to explain the {\it
Chandra} LETGS spectrum of \kpd\ and the energetic impossibility of a
coronal explanation, there are two additional puzzles regarding the
spectrum of this object at both shorter and longer wavelengths that
cannot be explained by our photospheric models.  These are the
detection of flux at $\sim 1$~keV (12.4~\AA ) by
\citet{O'Dwyer.etal:03}, and the observation of $\Delta n=1$ emission
lines of O~VIII between levels $n=7$-10 at UV and optical wavelengths
\citep{Werner.etal:94,Sion.Downes:92,Sion.etal:97}.  It seems likely
that the two are related.

We have verified that there is no trace of any flux near 12~\AA\ in
the LETGS spectrum, although the flux density reported by
\citet{O'Dwyer.etal:03} based on ROSAT observations (5 count~s$^{-1}$
in the Boron filter) lies below our
LETGS detection limit in this short observation: only $\sim 18$ LETG+HRC-S
counts are expected, spread over a fairly large detector area over
which the background is an order of magnitude larger.

More puzzling is the lack of O~VIII lines in the {\it Chandra} data:
the excited higher $n$ states will decay through cascades to lower
$n$, and we would expect to see some fraction of this decay channel in
the Lyman series resonance lines.  Based on the Goddard High
Resolution Spectrograph UV spectrum of \kpd\ obtained on 1994 June 1
and analysed by \citet{Werner.etal:96}, the O~VIII 2976.57~\AA\
$n=8$-7 transition flux is 0.07~ph~cm$^{-2}$~s$^{-1}$.  The upper
limit to the flux of the O~VIII 18.98~\AA\ $n=2$-1 doublet in the {\it
Chandra} spectrum is more than three orders of magnitude less than
this.  

There are two possible explanations for the absence of the
O~VIII Ly$\alpha$ lines: (1) the source is variable and during the
{\it Chandra} observation the O~VIII lines were much weaker; (2) the
X-ray lines are suppressed by some mechanism.  While the lack of any
simultaneous observations accompanying the {\it Chandra} pointing
prevents us drawing definitive conclusions regarding (1), we note that
optical spectra of \kpd\ similar to those described by
\citet{Werner.etal:94} and obtained at the Calar Alto 3.5m telescope
in 1991 July, 1992 September and 1994 May, exhibit identical O~VIII
$n=10$-9 and 9-8 lines and conclude that large amplitude variability
in O~VIII is unlikely.  Instead, the most likely explanation for the
lack of prominent X-ray O~VIII lines is that they are suppressed.

We speculate that the O~VIII Ly$\alpha$ lines are formed in a low
density medium or wind surrounding the white dwarf, as has been
suggested by earlier workers
\citep{Sion.Downes:92,Sion.etal:97,Werner.etal:94}.  In this case, the
low $n$ resonance lines might be sufficiently optically thick to
resonance scattering that line photons are destroyed by photoelectric
absorption as they undergo multiple scattering events within the
emitting region.  The line centre optical depth, $\tau$, in a
He-dominated low-density medium can be written
\citep[e.g.,][]{Acton:78,Mariska:92}:
\begin{equation}
 \tau = 1.16 \cdot 10^{-14} \cdot
     \frac{n_{\mathrm{i}}}{n_{\mathrm{el}}} A_{\mathrm{Z}}
    \frac{n_{\mathrm{H}}}{n_{\mathrm{e}}}
    \lambda f \sqrt{\frac{M}{T}} n_{\mathrm{e}} \ell 
    \label{e:tau}
\end{equation}
for ion fraction ($n_{\mathrm{i}}/n_{\mathrm{el}}$,
element abundance $A_{\mathrm{Z}}$, oscillator strength $f$,
temperature $T$, electron density $n_{\mathrm{e}}$, atomic weight $M$,
and where $n_{\mathrm{He}}/n_{\mathrm{e}} \sim 0.5$, 
and $\ell$ is the total path length along the line of sight
through the emitting plasma.  

Taking O~VIII 18.97~\AA\ $2p\rightarrow 1s$ as an example, we have
estimated the optical depth for two limiting cases of the morphology
of the emitting plasma: (a) a ``corona'' in hydrostatic equilibrium,
for which the scale height, or typical pathlength, $\ell$, is much
smaller than the stellar radius, $\ell << R_\star$; (b) an extended
wind for which $\ell >> R_\star$.  We have used PINTofALE
\citep{Kashyap.Drake:00} to estimate the total volume emission
measure, $n_e^2V$, of the O~VIII emitting plasma based on the total of
25 ROSAT (Boron filter) counts observed near 1~keV in 5~ks by
\citet{O'Dwyer.etal:03}.  We assumed that the plasma is isothermal
with $T=2\times 10^6$~K (corresponding to the peak in the O$^{7+}$ ion
fraction, such that $n_{\mathrm{i}}/n_{\mathrm{el}}\sim 1$), is
collision-dominated, has a He-dominated composition as listed in
Table~1 and is absorbed by an ISM column density
represented by a neutral H column of $N_H=5\times 10^{20}$~cm$^{-2}$.
For the approximate distance of 270~pc estimated by
\citet{Werner.etal:04}, the observed ROSAT counts imply a volume
emission measure $n_e^2V\sim 10^{54}$~cm$^{-3}$.

Adopting the stellar parameters of \kpd\ from \citet{Werner.etal:94}, 
the pressure scale height for a He plasma in hydrostatic equilibrium,
$kT/m_{He}g_\star=6.2\times 10^6$~cm, or 0.2~\%\ of the stellar
radius and the plasma density is $\sim 10^{14}$~cm$^{-3}$.  Based on
Equation~\ref{e:tau}, the optical depth in O~VIII Ly$\alpha$ is
$\tau=17$---ie extremely large.  For a much more extended plasma, the
line centre optical depth is $\tau\sim 200$ for $\ell \sim 10 R_\star$
and remains very large ($\tau > 10$) for emitting size scales up to
$10^4R_\star$. 

The large optical depths in the O~VIII Ly$\alpha$ resonance line imply
that these line photons are trapped within the plasma and must undergo
many scattering events before escaping.  If the plasma has a
temperature of $\sim 2\times 10^6$~K, it will contain a significant
fraction of H-like and He-like ions among the light elements C, N and
O, together with ions with $n=3$ ground states of elements such as Fe
if these are present.  The total photoelectric absorption
cross-section for our postulated plasma, computed using the
Hartree-Dirac-Slater photoionisation cross-sections of
\citet{Verner.etal:93} and \citet{Verner.Yakovlev:95}, together with
ion fractions from \citet{Mazzotta.etal:98}, amounts to $\sigma\sim
10^{-23}$~cm$^2$.  For the plasma radial extents considered above, for
which O~VIII Ly$\alpha$ is optically thick, the typical column density
of the plasma in terms of He nuclei ranges from several $10^{20}$ to
$10^{22}$~cm$^{-2}$.  Absorption for weaker, optically thin lines and
continuum is therefore negligible.  However, the trapped O~VIII line
photons must undergo such a large number of scattering events that
they are destroyed by photoelectric absorption before they can escape.
Similar arguments apply to higher Lyman series lines; while their
f-values are smaller, they are also effectively quenched through
decays to levels $n> 1$ and then eventually through Ly$\alpha$ as soon
as scattering depths become significant.  O~VIII 16.01~\AA\ Ly$\beta$,
for example, is lost to resonant scattering in decays through the
$n=3$-2 Balmer line at 102.43~\AA , and then through Ly$\alpha$.
Unfortunately, the ISM absorbing column to \kpd\ is too large for the
Balmer lines to be visible, though a future detection of O~VIII
Ba$\alpha$ in other similar objects to \kpd\ would provide a useful
test of this model.

If the $\sim 1$~keV emission found by \citet{O'Dwyer.etal:03} is
indeed due to a wind or outflow, then an optically-thin plasma model
with a temperature of $2\times 10^6$~K accounting for the observed
ROSAT counts at this energy has a total luminosity of $\sim 5\times
10^{31}$~erg~s$^{-1}$, or $10^{-4}$ times the stellar bolometric
luminosity.  The 0.1-2.5~keV X-ray to bolometric luminosity ratio is
$\sim 2\times 10^{-5}$---a somewhat higher value than observed for OB
stars, whose X-rays are believed to originate in shocked winds and 
for which the X-ray to bolometric luminosity ratio
$L_X/L_{bol}\sim 10^{-6}$-$10^{-8}$ \citep[e.g.][]{Berghoefer.etal:97}.

\section{Conclusions}

We have shown that a coronal plasma is unable to match the {\it
Chandra} LETG+HRC-S 20-80~\AA\ spectrum of \kpd\ and that coronal
models are energetically implausible as an origin for this observed
soft X-ray flux.  This part of the soft
X-ray spectrum of \kpd\ can instead be explained by photospheric
models containing trace amounts of heavier elements.  Photospheric
models do not, however, explain the soft X-ray emission at shorter
wavelengths ($\sim 12$~\AA ; 1 keV) revealed recently by
\citet{O'Dwyer.etal:03} and discussed in more detail by
\citet{Chu.etal:04b}.  The origin of this emission remains mysterious,
though an outflow or wind with a temperature of $\sim 2\times 10^6$~K
is able to explain both the X-rays and the presence of high $n$ O~VIII
UV-optical emission lines but an absence of significant O~VIII
Ly$\alpha$ flux in the {\it Chandra} spectrum.  
In this scenario, the Ly$\alpha$ photons are trapped
by resonance scattering and destroyed by photoelectric absorption.  If
an outflow is responsible for the 1~keV X-rays, the total luminosity
of this plasma amounts to $\sim 10^{-4}L_{bol}$, which is a factor of
100 or so larger than for winds of OB stars.

\acknowledgments

We thank the NASA AISRP for providing financial assistance for the
development of the PINTofALE package, and the CHIANTI project for
making publicly available the results of their substantial effort in
assembling atomic data useful for coronal plasma analysis.  Brad
Wargelin is thanked for useful comments.  JJD was
supported by NASA contract NAS8-39073 to the {\em Chandra X-ray
Center} during the course of this research. X-ray data analysis 
in T\"ubingen is supported by the DLR under grant 50\,OR\,0201.




\begin{thebibliography}{26}
\expandafter\ifx\csname natexlab\endcsname\relax\def\natexlab#1{#1}\fi

\bibitem[{{Acton}(1978)}]{Acton:78}
{Acton}, L.~W. 1978, \apj, 225, 1069

\bibitem[{{Barstow} {et~al.}(1995){Barstow}, {Holberg}, {Werner}, \&
  {Nousek}}]{Barstow.etal:95}
{Barstow}, M.~A., {Holberg}, J.~B., {Werner}, K., \& {Nousek}, J.~A. 1995,
  Advances in Space Research, 16, 73

\bibitem[{{Berghoefer} {et~al.}(1997){Berghoefer}, {Schmitt}, {Danner}, \&
  {Cassinelli}}]{Berghoefer.etal:97}
{Berghoefer}, T.~W., {Schmitt}, J.~H.~M.~M., {Danner}, R., \& {Cassinelli},
  J.~P. 1997, \aap, 322, 167

\bibitem[{{Cassinelli}(1982)}]{Cassinelli:82}
{Cassinelli}, J.~P. 1982, Advances in Space Research, 2, 67

\bibitem[{{Chayer} {et~al.}(1995){Chayer}, {Fontaine}, \&
  {Wesemael}}]{Chayer.etal:95}
{Chayer}, P., {Fontaine}, G., \& {Wesemael}, F. 1995, \apjs, 99, 189

\bibitem[{{Chu} {et~al.}(2004){Chu}, {Gruendl}, {Williams}, {Gull}, \&
  {Werner}}]{Chu.etal:04b}
{Chu}, Y., {Gruendl}, R.~A., {Williams}, R.~M., {Gull}, T.~R., \& {Werner}, K.
  2004, \aj, 128, 2357

\bibitem[{{Downes} {et~al.}(1985){Downes}, {Liebert}, \&
  {Margon}}]{Downes.etal:85}
{Downes}, R.~A., {Liebert}, J., \& {Margon}, B. 1985, \apj, 290, 321

\bibitem[{{Fleming} {et~al.}(1993){Fleming}, {Werner}, \&
  {Barstow}}]{Fleming.etal:93}
{Fleming}, T.~A., {Werner}, K., \& {Barstow}, M.~A. 1993, \apjl, 416, L79+

\bibitem[{{Kashyap} \& {Drake}(2000)}]{Kashyap.Drake:00}
{Kashyap}, V.~L. \& {Drake}, J.~J. 2000, Bulletin of the American Astronomical
  Society, 32, 1227

\bibitem[{{Mariska}(1992)}]{Mariska:92}
{Mariska}, J.~T. 1992, {The solar transition region} (Cambridge Astrophysics
  Series, New York: Cambridge University Press, |c1992)

\bibitem[{{Mazzotta} {et~al.}(1998){Mazzotta}, {Mazzitelli}, {Colafrancesco},
  \& {Vittorio}}]{Mazzotta.etal:98}
{Mazzotta}, P., {Mazzitelli}, G., {Colafrancesco}, S., \& {Vittorio}, N. 1998,
  \aaps, 133, 403

\bibitem[{{Miksa} {et~al.}(2002){Miksa}, {Deetjen}, {Dreizler}, {Kruk},
  {Rauch}, \& {Werner}}]{Miksa.etal:02}
{Miksa}, S., {Deetjen}, J.~L., {Dreizler}, S., {Kruk}, J.~W., {Rauch}, T., \&
  {Werner}, K. 2002, \aap, 389, 953

\bibitem[{{Miller}(2002)}]{Miller:02}
{Miller}, N.~A. 2002, in ASP Conf. Ser. 277: Stellar Coronae in the Chandra and
  XMM-NEWTON Era, ed. F.~{Favata} \& J.~J. {Drake}, San Francisco, 379

\bibitem[{{O'Dwyer} {et~al.}(2003){O'Dwyer}, {Chu}, {Gruendl}, {Guerrero}, \&
  {Webbink}}]{O'Dwyer.etal:03}
{O'Dwyer}, I.~J., {Chu}, Y., {Gruendl}, R.~A., {Guerrero}, M.~A., \& {Webbink},
  R.~F. 2003, \aj, 125, 2239

\bibitem[{{Otte} {et~al.}(2004){Otte}, {Dixon}, \& {Sankrit}}]{Otte.etal:04}
{Otte}, B., {Dixon}, W.~V.~D., \& {Sankrit}, R. 2004, \apjl, 606, L143

\bibitem[{{Schuh} {et~al.}(2002){Schuh}, {Dreizler}, \&
  {Wolff}}]{Schuh.etal:02}
{Schuh}, S.~L., {Dreizler}, S., \& {Wolff}, B. 2002, \aap, 382, 164

\bibitem[{{Sion} \& {Downes}(1992)}]{Sion.Downes:92}
{Sion}, E.~M. \& {Downes}, R.~A. 1992, \apjl, 396, L79

\bibitem[{{Sion} {et~al.}(1997){Sion}, {Holberg}, {Barstow}, \&
  {Scheible}}]{Sion.etal:97}
{Sion}, E.~M., {Holberg}, J.~B., {Barstow}, M.~A., \& {Scheible}, M.~P. 1997,
  \aj, 113, 364

\bibitem[{{Vaiana} {et~al.}(1981){Vaiana}, {Cassinelli}, {Fabbiano},
  {Giacconi}, {Golub}, {Gorenstein}, {Haisch}, {Harnden}, {Johnson}, {Linsky},
  {Maxson}, {Mewe}, {Rosner}, {Seward}, {Topka}, \& {Zwaan}}]{Vaiana.etal:81}
{Vaiana}, G.~S., {Cassinelli}, J.~P., {Fabbiano}, G., {Giacconi}, R., {Golub},
  L., {Gorenstein}, P., {Haisch}, B.~M., {Harnden}, F.~R., {Johnson}, H.~M.,
  {Linsky}, J.~L., {Maxson}, C.~W., {Mewe}, R., {Rosner}, R., {Seward}, F.,
  {Topka}, K., \& {Zwaan}, C. 1981, \apj, 245, 163

\bibitem[{{Verner} \& {Yakovlev}(1995)}]{Verner.Yakovlev:95}
{Verner}, D.~A. \& {Yakovlev}, D.~G. 1995, \aaps, 109, 125

\bibitem[{{Verner} {et~al.}(1993){Verner}, {Yakovlev}, {Band}, \&
  {Trzhaskovskaya}}]{Verner.etal:93}
{Verner}, D.~A., {Yakovlev}, D.~G., {Band}, I.~M., \& {Trzhaskovskaya}, M.~B.
  1993, Atomic Data and Nuclear Data Tables, 55, 233

\bibitem[{{Werner} {et~al.}(2003){Werner}, {Deetjen}, {Dreizler}, {Nagel},
  {Rauch}, \& {Schuh}}]{Werner.etal:03}
{Werner}, K., {Deetjen}, J.~L., {Dreizler}, S., {Nagel}, T., {Rauch}, T., \&
  {Schuh}, S.~L. 2003, in ASP Conf. Ser. 288: Stellar Atmosphere Modeling,
  31--1

\bibitem[{{Werner} {et~al.}(1996){Werner}, {Dreizler}, {Heber}, {Rauch},
  {Fleming}, {Sion}, \& {Vauclair}}]{Werner.etal:96}
{Werner}, K., {Dreizler}, S., {Heber}, U., {Rauch}, T., {Fleming}, T.~A.,
  {Sion}, E.~M., \& {Vauclair}, G. 1996, \aap, 307, 860

\bibitem[{{Werner} {et~al.}(1994){Werner}, {Heber}, \&
  {Fleming}}]{Werner.etal:94}
{Werner}, K., {Heber}, U., \& {Fleming}, T. 1994, \aap, 284, 907

\bibitem[{{Werner} {et~al.}(2004){Werner}, {Rauch}, {Barstow}, \&
  {Kruk}}]{Werner.etal:04}
{Werner}, K., {Rauch}, T., {Barstow}, M.~A., \& {Kruk}, J.~W. 2004, \aap, 421,
  1169

\bibitem[{{Young} {et~al.}(2003){Young}, {Del Zanna}, {Landi}, {Dere}, {Mason},
  \& {Landini}}]{Young.etal:03}
{Young}, P.~R., {Del Zanna}, G., {Landi}, E., {Dere}, K.~P., {Mason}, H.~E., \&
  {Landini}, M. 2003, \apjs, 144, 135

\end{thebibliography}



\end{document}